\documentclass[12pt,dvips]{article}
\usepackage{epsfig}

\begin{document}
%%%%%%%%%%%%%%%%%%%%%%%%%%%%%%%%%%%%%%%%%%%%%%%%%%%%%%%%%%%%%%%%%%%%%%%%%%%
\title{
\vspace{-5.0cm}
\begin{flushright}
{\normalsize UNIGRAZ-}\\
\vspace{-0.3cm}
{\normalsize UTP-}\\
\vspace{-0.3cm}
{\normalsize 11-06-96}\\
\end{flushright}
\vspace*{1cm}
%\vfill
Four-dimensional pure compact U(1) gauge theory on a
spherical lattice}

\author
{\bf J.~Jers{\'a}k\vspace{12pt} \\
Institut f{\"u}r
Theoretische Physik E,\\RWTH Aachen, Germany\vspace{12pt} \\
\bf C.~B.~Lang\\ \\
Institut f{\"u}r Theoretische Physik, \\
Karl-Franzens-Universit\"at Graz, Austria\vspace{12pt} \\
\bf T.~Neuhaus\\ \\
FB8 Physik, BUGH Wuppertal, Germany}
\date{\today}
\maketitle
\thispagestyle{empty}

\begin{abstract}
We investigate the confinement-Coulomb phase transition in the
four-dimensional (4D) pure compact U(1) gauge theory on spherical
lattices. The action contains the Wilson coupling $\beta$ and the
double charge coupling $\gamma$. The lattice is obtained from the 4D
surface of the 5D cubic lattice by its radial projection onto a 4D
sphere, and made homogeneous by means of appropriate weight factors for
individual plaquette contributions to the action. On such lattices the
two-state signal, impeding the studies of this theory on toroidal
lattices, is absent for $\gamma \le 0$. Furthermore, here a consistent
finite-size scaling behavior of several bulk observables is found, with
the correlation length exponent $\nu$ in the range $\nu = 0.35 - 40$.
These observables include Fisher zeros, specific-heat and cumulant
extrema as well as pseudocritical values of $\beta$ at fixed $\gamma$.
The most reliable determination of $\nu$ by means of the Fisher zeros
gives $\nu = 0.365(8)$. The phase transition at $\gamma \le 0$ is thus
very probably of 2$^{\rm nd}$ order and belongs to the universality
class of a non-Gaussian fixed point.
\end{abstract}

%%%%%%%%%%%%%%%%%%%%%%%%%%%%%%%%%%%%%%%%%%%%%%%%%%%%%%%%%%%%%%%%%%%%%%%%
\section{Introduction}
%%%%%%%%%%%%%%%%%%%%%%%%%%%%%%%%%%%%%%%%%%%%%%%%%%%%%%%%%%%%%%%%%%%%%%%%

\subsection{Motivation}

The introduction of a space-time lattice for quantum field theories
serves several purposes. It provides a regularization for the
renormalization scheme; it allows us to apply efficient computational
methods to perform the functional integrals; it may be considered a
mere approximation scheme for the problem in the continuum. Among the
four-dimensional (4D) gauge field theories with Lie groups the one
with $U(1)$ symmetry at first sight appears to be the simplest to
test this approach.  It is also of considerable
interest as it has many features in common with QCD, like a confining
strong coupling phase, topological excitations and gauge-balls. In
addition it shows a phase transition (PT) to a phase with weak
coupling signature, a massless photon and long range interaction. In
fact it has been the first lattice gauge model with continuous gauge
group to be studied by the computational methods that became available
in the 80's \cite{CrJa79b}.

Below (in subsect. \ref{SubsectionSituation}) we will discuss the
various results obtained since. However, in summary we may say that up
to now there is no definite answer to the critical properties of its
PT. In most simulations a two-state signal at the PT
indicated a $1^{\rm st}$ order transition. On the other hand, the
critical behavior according to such a transition has not been confirmed
in thorough finite-size scaling (FSS) studies.  The problem persisted
when the original Wilson action containing only $\cos(\Theta_P)$ was
extended to include the double charge coupling
\begin{equation}\label{ACTION}
S = -\sum_P \left( \beta \cos(\Theta_P) + \gamma \cos(2\Theta_P)\right) .
\end{equation}
Here $\Theta_P \in [0,2\pi)$ is the plaquette angle, i.e.  the argument
of the product of $U(1)$ link variables around a plaquette $P$.  It was
conjectured, that the $1^{\rm st}$ order PT changes into a $2^{\rm nd}$
order one at a tricritical point (TCP) at small negative values of
$\gamma$, but that was never confirmed in actual simulations at $\gamma
\le 0$.  If there is indeed  a $2^{\rm nd}$ order transition, its
properties have not been determined up today.

On the other hand, both computational and data ana\-ly\-sis techniques
have progressed. This provides us with the possibility to perform a
thorough FSS study of this model in a new context. Practically all
other studies have dealt with the standard periodic boundary
conditions, i.e. hypertorus topology for the lattices. It has however
been realized that there are nonlocal excitations in the system ---
closed monopole loops --- that may extend over the whole lattice.
Therefore the essentially local updating algorithms used for gauge
theories, together with the boundary conditions, may affect
thermalization properties. One expects, that in the thermodynamic limit
the ``continuum'' properties of the system are independent of the
global topology of the system, if this becomes locally flat. For these
reasons it was suggested to simulate the model on lattices with
spherelike topology \cite{LaNe}, amounting to modified boundary
conditions, such that closed loops are always homotopically equivalent
to points.  The spherelike topology allows the monopoles more freedom
in their dynamics without changing the action.

One generally expects that the thermodynamic properties of the bulk
phase (defined by the behavior of the free energy per unit volume in
the thermodynamic limit) are not affected by contributions which grow
slower than the total volume.  Boundary contributions are suppressed
$O(1/L)$ relative to the leading term, curvature terms at least
$O(1/L^2)$, and therefore they should not change the critical exponents
of the bulk phase.  This does not necessarily hold for the ground
state. E.g. at phase transitions of $1^{\rm st}$ order the phase
mixture may be different, depending on boundaries or even one
individual spin, or due to an overall external field vanishing
$O(1/L^D)$. A similar influence may come from the curvature, even if
thinned out over the volume in $O(1/L^2)$.  However, if the manifold
becomes locally flat in the thermodynamic limit the systems universal
{\em critical} properties should be independent of the global
topological structure. Otherwise we could hardly assume that we can do
reliable physics on earth without definite knowledge about the
topological details of the universe.

Whereas in the original study \cite{LaNe} the surface of a 5D cube
was used, we here choose a discretization of the sphere, where the
curvature is distributed more homogeneously over the lattice.  Although
the system is nonhomogeneous on the scale of the lattice constant, it
is homogeneous and isotropic on larger scales.  As will be demonstrated
here, in this 4D system with the topology of the surface of a 5D sphere
we find no two-state signal on lattices with up to almost 20$^4$
points.  Of course we cannot exclude the possibility, that a two-state
signal reappears for even larger lattices.  However, our FSS analysis
leads to consistent results for a PT of $2^{\rm nd}$ order for
$\gamma\leq 0$.

The best measurement of the correlation length critical exponent $\nu$
by means of the FSS behavior of the Fisher zero gives $\nu=0.365(8)$.
Less precise FSS analyses of several other bulk observables are
consistent with $\nu$ values in the interval $\nu = 0.35 - 0.40$. As
we argue in \cite{JeLa96a}, due to rigorous dual
relationships our results imply that also the Coulomb gas of monopole
loops\cite{BaMy77} and the noncompact U(1) Higgs model at large
negative squared bare mass (frozen superconductor)\cite{Pe78} have a
continuum limit described by the same non-Gaussian fixed point. Some
further related models are discussed in \cite{JaKl95}.

The first order signal persists --- also for the discussed spherelike
geometries --- at values $\gamma\simeq 0.2$. Since scaling may be
garbled close to TCPs we concentrated on negative values of the double
charge coupling $\gamma$ in our study. Scaling and FSS is expected to
improve at larger distance from the  $1^{\rm st}$ order part of the PT
line. Nevertheless, at $\gamma = 0$ the two-state signal is still
absent and the scaling behavior is consistent with that found at
$\gamma < 0$.

Let us add a remark on the extended action considered.  Some time ago
it was pointed out \cite{Gr81}, that, although the Wilson-  and  the
heat kernel (Villain) action do have reflection positivity, some
actions do not. Positivity is a sufficient, but not a necessary
condition for unitarity; it guarantees the existence of a positive
definite  scalar product and the spectral condition, one of the formal
conditions for the existence of a continuum limit field theory
\cite{OsSe78}. The actions with the parameter values $\gamma<0$
considered here are not reflection positive. They share this property
with other actions like e.g. the (Symanzik) improved actions.

On the other hand, if reflection positivity holds on a part of a
critical surface that is in the domain of attraction of a fixed point
of some renormalization group transformation, we expect that it should
be fulfilled everywhere in that domain. Unitarity violating states
like ghosts should then decouple at large scale.  We therefore find it
justified to study the action near candidates for critical points even
outside the region $\gamma \ge 0$, where reflection positivity
is respected on the scale of the lattice spacing.  Unitarity at
$\gamma < 0$ is also suggested by the observation that the regions
with $\gamma < 0$ and $\gamma \ge 0 $ are connected
by the RG flows \cite{La86}.

Following a brief review of the situation in the U(1) pure gauge model
we then introduce the spherical lattice in sect.
\ref{SectionLatticeGeometry}.  (Further technical details are given in
the appendix.) In sect. \ref{SectionMethods} we present the Monte Carlo
simulation and discuss the observables in some detail, including the
expected FSS behavior. The results and their analysis are summarized in
sect. \ref{SectionResults}, followed by our conclusions.

\subsection{Situation of U(1) pure gauge studies}
\label{SubsectionSituation}

U(1) is the most elementary Lie group that can be used to construct a
quantum gauge field theory. Yet, when formulated on a lattice, the pure
U(1) gauge theory turned out unexpectedly to be one of the most
intriguing and less understood quantum gauge models. The awareness of
this fact has developed with an accumulation of the numerical
experience.  In this section we give a brief description of this
historical development. We apologize for quoting only a subjectively
chosen part of a much larger number of valuable papers.

Since the introduction of lattice gauge theories by Wilson
\cite{Wi74}, the pure compact U(1) theory has been of interest as a
theory with a rigorously established \cite{Gu80,FrSp82} PT between the
confinement and the free charge (Coulomb) phases at zero temperature.
One reason was the importance of topological excitations, the
monopoles, for confinement, as manifested by their prominent r{\^o}le
in this phase transition \cite{BaMy77,Po75,Sa77,GlJa77,DeTo80,BaShr}.
Another purpose was to study it as a prototype example for applications
of numerical methods of statistical physics in a lattice gauge theory,
in particular an investigation of the continuum limit at the phase
transition. However, the lesson has been that this phase transition
provides no simple exercise.

In the very first numerical investigations
\cite{CrJa79b,LaNa80,Bh81,DeTo81,Mo82}, restricted to $\gamma = 0$ and
small lattices, a behavior consistent with a $2^{\rm nd}$ order phase
transition at $\beta \simeq 1$ was observed. But this order was
questioned by the subsequent observation of a two-state signal on
larger lattices \cite{JeNe83}. Such a signal could imply that the phase
transition at $\gamma = 0$ is actually of weak $1^{\rm st}$ order,
which would prevent taking a continuum limit there. The question was,
and remained to be, whether this signal may be a finite-size effect.

In the model with extended Wilson action (\ref{ACTION}) it was found
\cite{Bh82} that the confinement-Coulomb phase transition is clearly of
$1^{\rm st}$ order for $\gamma \ge 0.2$, and weakens with decreasing
$\gamma$. This suggested that the order of the transition changes when
$\gamma$ is decreased, presumably at a TCP.

The question at which value of $\gamma$ this happens turned out to be
very difficult. First, even at large negative $\gamma$, a two-state
signal has been observed, e.g. at $\gamma = -0.5$ on the $8^4$ lattice
\cite{EvJe85}. Second, TCPs are known to cause intricate finite-size
effects \cite{LaSw81}, easily mocking up a false order of the phase
transition.

In the hope to clarify the situation, an investigation of the strongly
$1^{\rm st}$ order part of the phase transition line at $\gamma \ge
0.2$ was performed \cite{EvJe85}. There the latent heat $\Delta e$
can be determined reliably even on moderately large
lattices. Its independence on the lattice size was checked very
carefully. The extrapolation of $\Delta e$ to zero with
decreasing $\gamma$ by means of the power law
\begin{equation}
               \Delta e \propto (\gamma -\gamma^{TCP})^{\beta_u},
        \label{POWERLAW}
\end{equation}
suggested that the order of the transition changes at the TCP with
$\gamma^{TCP} = -0.11(5)$, implying a $1^{\rm st}$ order PT at $\gamma
= 0$.

This extrapolation procedure is an attempt to control finite-size
effects, but it uses the assumption that the power law behavior
(\ref{POWERLAW}), which the data in the investigated region $\gamma =
0.2 - 0.5$ are consistent with, indeed holds throughout the whole
interval between $\gamma =0.5$  and $\gamma = \gamma^{TCP}$.  This
assumption has remained unverified.  Another possible uncertainty in
\cite{EvJe85} was the determination of $\Delta e$ at a strong $1^{\rm
st}$ order transition, without the more advanced methods of
investigation of such transitions \cite{BeNe}.

Monte Carlo RG (MCRG) studies \cite{Bu86,GuNo86,La86,Ha88} did not
confirm this position of the TCP at negative $\gamma$. Of course, also
the MCRG approach suffers from ambiguities due to a small number of RG
steps and a restricted number of couplings considered. Therefore also
these studies remained inconclusive about the order of the transition
around $\gamma = 0$ in the thermodynamic limit (although they all
observed clear two-state signals).

In spite of this, numerous attempts to determine the critical exponent
$\nu$ provided roughly consistent values in the range $\nu \simeq 0.28
- 0.42$. These studies used various methods: the analytic calculations
\cite{Ha81,IrHa84}, the FSS analysis \cite{LaNa80,Bh81,Bh82,MuSc82},
the scaling of the string tension \cite{DeTo81,Mo83,Ca83,JeNe85}, and
the MCRG method \cite{Bu86,GuNo86,La86,LaRe87,Ha88}.  Three actions:
Wilson--, extended Wilson--, and Villain--type have been used.

This suggested that the pure compact U(1) lattice theory might have an
interesting continuum limit at the confinement-Coulomb PT, pondered,
e.g., in \cite{Pe78,Cr81}.  However, the two-state signal, observed on
finite lattices even for $\gamma < 0$ \cite{EvJe85} as well as for the
Villain action \cite{LaRe87}, hindered the investigations of this
possibility. Even if this signal is only a finite-size effect and the
transition in the infinite volume limit is genuinely of $2^{\rm nd}$
order, it represents a serious impediment for a precise FSS analysis or
MCRG studies. Because of this the investigation of the pure compact
U(1) gauge theory lost its momentum.  Until now there is no established
$2^{\rm nd}$ order PT with an undisputed determination of critical
indices in this model.

All the above mentioned numerical work has been performed on 4D
toroidal lattices. Recently, following earlier suggestions
\cite{GrJa85,GuNo86,La86}, two of the present authors speculated that
the two-state signal at $\gamma \le 0$ may be related to monopole loops
winding around the toroidal lattice, and trapped in simulations with
local update algorithms \cite{LaNe}. They used the 4D surface of a 5D
cubic lattice instead of the torus, and observed that at $\gamma = 0$
the two-state signal vanishes on lattices of all investigated sizes.

Choosing a spherelike topology provides a way to allow the monopoles
more freedom in their dynamics without affecting them locally by
changing the action. We consider this as preferable to adding terms to
the action that forbid or restrict the occurrence of monopoles produces
$O(L^D)$ contributions to the total free energy and thus changes the
bulk properties of the system \cite{BoMi93a,KeRe94}. In that case one
explores the phase diagram in different regions of the space of
couplings and the position of the phase transition in $\beta$ moves to
different values, depending on the extra couplings. None of these
studies has led to phase transitions of $2^{\rm nd}$ order, though.

However, the cause for the two-state signal on the toroidal lattice,
and of its vanishing on a lattice with the topology of a sphere, is not
yet fully understood. Possibly the trivial $1^{\rm st}$ homotopy group
of such a lattice allows a smooth vanishing of winding monopole loops
in simulations. But some other recent results do not seem  to support
this interpretation \cite{KeRe94,KeRe,BoLi}.

On the other hand, for a study of the continuum limit on lattices with
the topology of a sphere, a complete understanding of the dynamics of
the two-state signal on a torus is not really necessary. What is
required is a construction of a spherelike lattice which is
homogeneous, in order to avoid the possibly related problems with the
FSS analysis, encountered in \cite{LaNe}. Achieving that in this paper,
we hope to give a new momentum to the investigation of the
confinement-Coulomb phase transition in the pure compact U(1) gauge
theory on the lattice. A construction of the continuum limit appears
now to  be feasible.

%%%%%%%%%%%%%%%%%%%%%%%%%%%%%%%%%%%%%%%%%%%%%%%%%%%%%%%%%%%%%%%%%%%%%%%%
\section{Spherelike lattices and lattice geometry}
\label{SectionLatticeGeometry}
%%%%%%%%%%%%%%%%%%%%%%%%%%%%%%%%%%%%%%%%%%%%%%%%%%%%%%%%%%%%%%%%%%%%%%%%

In an attempt to formulate the theory without modification of the
locally defined plaquette action and without forbidding or hindering
the dynamic evolution of monopole loops, the lattice topology  was
modified.  The usual periodic (or antiperiodic) boundary conditions
correspond to the topology of a 4D torus $T^4$ with the first homotopy
group $\mathbf Z^4$.  Closed loops (or networks of loops) cannot
necessarily be contracted to a point and the corresponding ground state
may be classified accordingly. The original motivation to divert from
this lattice structure was to study the possible influence of this
property on the dynamics of the phase transition.

In \cite{LaNe} it was therefore suggested to simulate and study the
model on a lattice with spherelike topology, in particular on the
surface SH[N] of a 5D hypercubic lattice of size $N^5$. This lattice
may be best visualized in analogy to the 2D surface of a 3D cubic
lattice. It may also be considered as a collection of 10 hypercubic
lattices of size $N^4$, glued together at their boundaries.  This
implies that one can expect the same critical coupling in the
thermodynamic limit as for the usual torus. This was indeed verified in
the Monte Carlo calculations \cite{LaNe}.  Details and parameters of
the geometry are listed in the appendix.

The so-defined lattice is locally flat, except at certain plaquettes
((D-2)-dimensional elements), where the curvature is concentrated, a
well-known property of Regge skeletons. The unusual features include
plaquettes bordering only three 3D cubes (instead of the usual 4),
links bordering less than 6 plaquettes and sites with less than 8
links.  These curvaturelike contributions as we might call them in the
absence of a strict theory in 4D are suppressed $O(1/N^2)$ relative to
the leading terms in the action.

In an attempt to distribute these local inhomogeneities more uniformly
over the lattice we introduced the ``almost smooth'' spherical lattice
S[N].  In the construction we project sites, links and
plaquettes of SH[N] (or its dual
SH'[N]) onto the surface of a concentric 4D sphere and
introduce weight factors similar to those used by \cite{ChFrLe82} in
their study of random triangulated lattices,
\begin{eqnarray} \label{ACTIONWeight}
S&=& - \sum_P\; w_P [ \beta \cos(\Theta_P) + \gamma \cos(2 \Theta_P)]
\nonumber\\
&&\mbox{with} \quad w_p = {A_P' \over A_P} .
\end{eqnarray}
Here, $A_P$ and $A_P'$  denote the areas of the corresponding plaquette
and its dual, respectively, of the projected lattice.

As discussed in \cite{ChFrLe82} in the situation of triangulated random
lattices, one has to distribute the total integration volume over all
contributions to the action, i.e. the plaquette terms in our case.
This may be done with the help of the dual lattice, where to each site,
link or plaquette there is an associated dual 4D cube, 3D cube or
plaquette.  The dual lattice sites in our situation are constructed
from the barycenters of the 4D cubes that have been projected to the
sphere $S^4$, followed by an adjusting projection of these points to
the sphere. Further reasoning according to \cite{ChFrLe82} leads to the
weight factors $w_P$ in (\ref{ACTIONWeight}). This choice is not
unique, but reproduces the usual continuum action $F_{\mu\nu}^2$ in the
naive continuum limit $g\to 0$ ($\beta\to\infty$)  and is thus
homogeneous in this limit. We study the system at finite $\beta$; there
slight distortions from the regular spherical surface are possible.
The value of $\beta_{crit}$ might be modified due to the weight factors
and thus does not necessarily agree with that of the torus or SH.

Technically we have to introduce some approximations. Usually the
plaquettes -- constructed via the projection of the sites to the sphere
-- will not be flat.  Plaquette areas are therefore determined from the
sum of two triangles. Also for this reason the curvature in this
formulation will not be perfectly uniformly distributed. In order to
achieve this, we would have to rely on a triangulated lattice. This
would imply a significant change of the action, which we wanted to
avoid. On the other hand we expect these effects to become irrelevant
in a situation with a large correlation length. The consistency of the
found FSS behavior justifies these simplifications.

The connectivity properties of SH and ${S\,}$ are identical. In the
computer programs the geometry is implemented with tables and the
weight factors  $w_P$ are precalculated. During the development of the
program and in the early stages of the analysis we also determined the
monopole positions (on the dual lattice) and reproduced them
graphically. We observed the expected properties:
\begin{itemize}
\item
The monopole loops were always closed;
\item the smallest loops had length 3 (corner plaquettes on the dual
lattice);
\item they fluctuated freely, appearing and disappearing without
noticeable correlation with positions close to corners.
\end{itemize}
This also served as a check of the consistency of the
connectivity tables.

%Insert table 1 here
\begin{table}
\caption{Effective volumes for the
studied lattices SH[N]. We also give the value of
$L = V^\frac{1}{4}$ which would give the
base length for a hypertorus lattice with the same volume.
\label{table1}}
\begin{center}
\begin{tabular}{rrr}
\hline
$N$ & $V$ & $L$ \\
\hline
\hline
 4     & 825.1   & 5.4\\
 5     & 2576.6  & 7.1 \\
 6     & 6268.1  & 8.9 \\
 7     & 12986.9  & 10.7 \\
 8     & 24064.1 & 12.5\\
 9     & 41074.6 & 14.2\\
10     & 65837.3 & 16.0 \\
12     & 147113.8 & 19.6\\
\hline
\end{tabular}
\end{center}
\end{table}

In our discussions we will refer to the {\em effective} lattice volume
\begin{equation}
V=\frac{1}{6}\sum_P\; w_P
\end{equation}
as the typical size quantity. For SH this would be just the number of
sites, for ${S\,}$ it is very close to this value. A length scale may
be defined as 
\begin{equation}
L\equiv V^\frac{1}{4}.
\end{equation}
Table \ref{table1} summarizes the effective volumes for the
lattice sizes used in our  study.

For strictly asymptotic dependencies as they come up in FSS studies it
is irrelevant, whether one uses $N$ or $L$. However, for moderately
sized finite systems a suitable choice improves the approach to the
asymptotic behavior.  Let us mention in this context that in
\cite{HoLa96} different lattice geometries were compared and it was
demonstrated, that the scaling curves show best agreement with each
other, if one indeed uses $L$ -- the size derived from the total
volume as opposed to the base length $N$ -- as size variable.  In the
present work we cannot compare with torus results, since for those the
two-state signal obscures the measured values of the cumulants.

%%%%%%%%%%%%%%%%%%%%%%%%%%%%%%%%%%%%%%%%%%%%%%%%%%%%%%%%%%%%%%%%%%%%%%%%
\section{Simulation methods and statistics}\label{SectionMethods}
%%%%%%%%%%%%%%%%%%%%%%%%%%%%%%%%%%%%%%%%%%%%%%%%%%%%%%%%%%%%%%%%%%%%%%%%

In an earlier study \cite{LaNe} we found that for lattice type SH
there are two-state signals at the pseudocritical $\beta$ for
$\gamma=0.2$, but no such indications at $\gamma=0$. For this reason we
now studied the action (\ref{ACTIONWeight}) of the spherical lattice
S at the values $\gamma = 0., -0.2, -0.5$.  Preliminary results have
been presented in \cite{JeLa95,HoJa96}.

\subsection{Updating and measuring}

We have worked with lattices S[N] for $N$ ranging between 4 and 12.
The couplings were chosen in the immediate neighborhood of the
pseudocritical values of $\beta$. For the analysis we determined the
histograms of the weighted sum of plaquette values
\begin{equation}
E \equiv \sum_P w_P \cos \Theta_P .
\end{equation}
Note that this is not the total energy, but just the part corresponding
to the coupling parameter $\beta$; it coincides with the total
plaquette energy for the Wilson action, $\gamma=0$.  Any scaling- or
two-state signal should be observable in that quantity.  We also define
the density
\begin{equation}
e \equiv E / \sum_P w_P = E / (6V) .
\end{equation}

We  combined the various histograms (for fixed $\gamma$ but different
$\beta$) with help of the Ferrenberg-Swendsen multihistogram
reweighting technique \cite{FeSw88}. For each $\gamma$ we thereby
construct the density of states $\rho(E;\gamma)$. The representation
\begin{equation} \label{PartFcn}
Z(\beta,\gamma) \equiv \sum_E \rho(E;\gamma) \exp{(-\beta E)}
\end{equation}
allows us to determine
various observables for continuous values of $\beta$ through
\begin{equation}
\langle E^n \rangle = \frac{1}{Z(\beta,\gamma)}
\sum_E \rho(E;\gamma) E^n \exp{(-\beta E)} .
\end{equation}
Since we never observed two-state signals we did not implement
multicanonical updating \cite{BeNe}.  We used a 3-hit
Metropolis update; for $\gamma=0$ we included an additional
overrelaxation step (the autocorrelation length decreased by a factor
of about 2).  For each lattice size at each $\gamma$ we
typically accumulated $O(10^6)$ updates, which is between $10^3$ and
$10^4$ times the integrated autocorrelation length for the observable
$E$ (cf. Table \ref{table2}).

The histograms had up to $10^4$ bins in order to exclude any possible
influence due to binning. In fact, by rebinning we found no changes
within single precision down to $O(500)$ bins. Due to the fine binning
the raw histograms have a noisy appearance, which is irrelevant for the
Ferrenberg-Swendsen analysis and for the representation
(\ref{PartFcn}). For the plots we use rebinned versions with only 500
bins maximum.

The total CPU-time spent for the calculations on workstations and on a
Cray-YMP sums up to 6800 hours in Cray-YMP units.

%Insert table 2 here
\begin{table}
\caption{Statistics of the data for the studied values of $\gamma$:
Lattice base length $N$, total number $n$ of configurations in multiples
of $10^6$, range of $\beta$-values, maximal $\tau_{int}$ values (from a
fit to all beta values, as discussed in the text; errors are typically
10\% of the values).
\label{table2}}
\begin{center}
\begin{tabular}{lrlr@{---}lr}
\hline
$\gamma$        &$N$&   $n/10^6$
&\multicolumn{2}{c}{$\beta$-range}  &$\tau_{int,max}$\\
\hline
\hline
0.              &4&     1.1&    0.98& 1.025    &12\\
                &6&     1.0&    1.& 1.025      &56\\
                &8&     1.1&    1.0125& 1.0275 &150\\
                &10&    1.47&   1.01& 1.025    &304\\
\hline
-0.2            &4&     1.1&    1.07& 1.2      &24\\
                &5&     1.8&    1.15& 1.175    &63\\
                &6&     1.6&    1.13& 1.21     &116\\
                &7&     1.1&    1.14& 1.175    &188\\
                &8&     1.6&    1.155& 1.175   &272\\
                &9&     1.1&    1.1655& 1.1715 &367\\
                &10&    1.0&    1.1635& 1.1715 &583\\
\hline
-0.5            &4&     4.8&    1.3& 1.65      &37\\
                &5&     1.8&    1.38& 1.47     &76\\
                &6&     2.1&    1.35& 1.5      &149\\
                &7&     1.9&    1.402& 1.452   &311\\
                &8&     1.5&    1.4& 1.455     &332\\
                &9&     1.8&    1.42& 1.442    &480\\
                &10&    1.55&   1.42& 1.442    &473\\
                &12&    1.6&    1.43& 1.44     &1565\\
\hline
\end{tabular}
\end{center}
\end{table}

\subsection{Observables and FSS}

\subsubsection{Cumulants}

We determined the $2^{\rm nd}$  and $4^{\rm th}$ order cumulants of the
observable $E$. Due to the analogy to the internal
energy (identical to $E$ only for $\gamma=0$)
we call the second order cumulant the specific-heat.
The specific-heat, the Challa-Landau-Binder
cumulant \cite{ChLaBi86} and another $4^{\mbox{\footnotesize th}}$
order cumulant suggested by Binder (cf. \cite{Bi92}
and \cite{FeLa91}) are defined through
\begin{eqnarray} \label{cumdefs}
c_V(\beta,L)&=& ~~\frac{1}{6V} \langle (E-\langle E\rangle )^2
\rangle  ,\\
V_{CLB}(\beta,L)&=& -\frac{1}{3} \frac{\langle (E^2-\langle
E^2\rangle )^2\rangle}{\langle E^2\rangle^2}  ,\\
U_4(\beta,L)&=& \frac{\langle (E-\langle E\rangle)^4\rangle}
{\langle(E-\langle E\rangle )^2\rangle^2}   .
\end{eqnarray}
The positions and values of their respective extrema are used for the
FSS analysis.

{}From the usual scaling hypothesis \cite{Fi72,FiBa72,Br82,Ca88,Ba83} one
expects for the singular part of the free energy density the scaling
behavior
\begin{equation}
f(\tau,L) = L^{-1/D} f(\tau L^{1/\nu}, 1) ,
\end{equation}
where $\tau \equiv (1-\beta/\beta_c)$ denotes the reduced coupling and
$L$ is a length scale parameter. From this one derives the scaling
behavior of the cumulants.  At a $2^{\rm nd}$ order phase transition
we expect (for D=4 and $\alpha>0$)
\begin{eqnarray}
c_{V,max}(L) &\simeq    &       L^{\alpha/\nu} \label{FSSrelCV}   ,\\
V_{CLB,min}(L) &\simeq  &       L^{\alpha/\nu - 4} \label{FSSrelVL}   ,\\
U_{4,min}(L)    &\simeq &       O(1) + O(L^{-\alpha/\nu})
\label{FSSrelU4}   ,\\
\beta_{c}(L)-\beta_c  &\simeq& L^{-\lambda} \label{FSSbeta}   .
\end{eqnarray}
For $\alpha=0$ there are logarithmic terms. The asymptotic value of
$U_{4,min}$ depends on the details of the distribution density $\rho(E)$ 
and is 3 for a Gaussian distribution. Mean field values are $\nu=1/2$ 
and with Josephson's law  $\alpha=2-D \nu=0$.

We denote by $\beta_c(L)$ definitions for pseudocritical points like
the positions of the extrema in the cumulants. The so-called
shift-exponent $\lambda$ is for many models equal to $1/\nu$, but not
necessarily so in general; such an identity is not a necessary result
of FSS (cf. the discussion in \cite{Ba83}). We return to this issue
later.  Furthermore, a priori we know nothing about the absolute size
of the multiplicative coefficients in the scaling formulas.
They depend on the details of the action, the lattice
geometry and the topology \cite{Ba83}.

For 1$^{\rm st}$ order transitions one expects the FSS  behavior
\begin{eqnarray}
L^{-D}  c_{V,max} &\rightarrow&  ~~\frac{1}{4}(e_o-e_d)^2  ,\\
V_{{CLB},min}  &
\rightarrow&
-\frac{1}{12}{(e_o^2-e_d^2)^2\over (e_oe_d)^2} + O(L^{-D}) ,\\
U_{4,min}  &\rightarrow&  1 +O(L^{-D}) ,
\\
\beta_c(L) - \beta_c  &=& O(L^{-D}) . \label{pscfss}
\end{eqnarray}
Here $e_o$ and $e_d$ denote the discontinuous values of the energy
density at the PT point.

As discussed, in the considered lattice geometry there are
inhomogeneities in the sense, that the coordination numbers of some
sites, links and plaquettes deviate from the usual torus numbers. For
the SH lattices these may be considered as lattice inhomogeneities.
Their contribution to the total free energy is suppressed $O(1/N^2)=
O(V^{-2/D})$. In our smoothed version of that lattice: S, the
inhomogeneous contribution should be smaller. There is however still a
possible contribution of the total curvature to the free energy, which
is suppressed with the same order (cf. also the discussion for the 2D
Ising model \cite{HoLa96}).  Thus --- in principle -- we also may
expect ``surface'' corrections of $O(V^{-2/D})$ in all FSS relations.
Indeed such contributions have been observed for the SH lattices
\cite{LaNe}. It turns out that they are much smaller in our present
study, in fact too small to study them.

\subsubsection{Fisher zeros\label{SecFisherZeros}}

Equation (\ref{PartFcn}) defines implicitly an analytic continuation to
complex values of $\beta$ not too far away from the real axis.
Therefore it is  possible to determine the nearby zeros of the
partition function \cite{YaLe52} in the complex $\beta$-plane, the
so-called Fisher zeros \cite{Fi68} (cf. \cite{FaMa81}).

One should add a warning concerning technical aspects.  The  histograms
are binned, having both, upper and lower limits for $E_{max}$ and
$E_{min}$ as well as a bin size $\Delta= (E_{max} - E_{min})\times
10^{-4}$. The representation (\ref{PartFcn}) for $\beta = \beta_R + i
\beta_I$ therefore is a discrete Fouriertransformation. It will induce
a periodicity in $\beta_I$ due to the bin size and an effective grid
with grid spacing $2 \pi/(E_{max} - E_{min})$ (although the values of
the partition function $Z$ are well-defined even between the grid
points, they carry no additional information).

Usually the distribution is similar to a Gaussian
distribution; let us for the sake of the argument assume such a form
\begin{equation}
\rho(E)\exp{(-\beta_R E)}  \simeq \exp{\left(-c (E-E_0)^2\right)} .
\end{equation}
{}From (\ref{PartFcn}) one then expects an oscillatory behavior of $Z$
proportional to $\exp{(i\beta_I E_0)}$. This is indeed observed in the
calculation.  In the search for partition function zeros one
starts with an identification of sign changes of $\mbox{Im}\,Z$ and
$\mbox{Re}\,Z$. The rapidly oscillating phase factor may confuse the
pattern and one has to work with a very fine resolution and to
carefully combine the sign-change analysis with a search in $\vert Z
\vert$. Also the grid structure may interfere with these oscillations
and one has to proceed with care.

So real and imaginary parts of the closest Fisher zeros provide further
(even) observables.  In particular the imaginary part of the zero
closest to the real axis provides a high quality estimator for the
critical exponent $\nu$ (cf. \cite{KeLa} for a recent high statistics
study of the 4D Ising model, where it was possible to identify the
logarithmic corrections to scaling on basis of the Lee-Yang
zeros \cite{YaLe52}). As will be demonstrated below this
quantity appears to have small corrections to the leading FSS behavior
in our environment; this observation is analogous to other recent
investigations of spherelike lattices in 2D \cite{GaLa95,HoLa96}.

{}From the scaling arguments for the free energy we expect
\begin{equation}
\vert z_0(L) - \beta_c\vert \simeq  L^{-1/\nu} \label{FSSFiZero} .\\
\end{equation}
This provides an upper bound for the real and imaginary part of $z_0$, in
particular
\begin{eqnarray}
\mbox{Im}\,z_0(L)& \simeq & O(L^{-1/\nu}) \label{FSSIm} ,\\
\mbox{Re}\,z_0(L) - \beta_c& \simeq & O(L^{-1/\nu})  .\label{FSSRe}
\end{eqnarray}
Although in some cases the angle, under which the zeros approach the
real axis (defined as the angle of a line connecting the two closest
zeros) is known (e.g. $\pi/2$ for the 2D Ising model in the Onsager
solution \cite{Fi68}, $\pi/4$ in the mean field solution for the 4D
$\phi^4$-model \cite{ItPe83}, both on cubic lattices with torus
topology) there is no FSS theory for this angle of approach with regard
to the size $L$.  Depending on details of the model, the geometry and
the topology, the {\em real part} --- which by analogy to the cumulants
we call a pseudocritical value --- may approach the asymptotic value
faster, i.e. with a shift exponent $\lambda$ larger than $1/\nu$
\cite{Ba83}.  Such a behavior has been observed in a recent study of
the 2D Ising model \cite{HoLa96}.

We also mention here, that the position of the closest
Fisher zero is related
to the peak position and value of the specific-heat. Since the partition
function may be expressed by the Vieta-product of all its
zeros $\{z_i\}$,  the specific-heat is proportional to
\begin{equation}
\sum_i \frac{1}{(\beta-z_i)^2}
\end{equation}
and therefore the leading contribution to $V\times c_V$ near the PT is
proportional to  $(\mbox{Im}\,z_0)^{-2}$. The peak position is
in leading order given by
$\mbox{Re}\,z_0$. Of course there are further contributions due to
the other zeros and a possible background from an entire function.

Also these observables may in principle exhibit corrections to FSS due
to curvature and topology as discussed above.

%%%%%%%%%%%%%%%%%%%%%%%%%%%%%%%%%%%%%%%%%%%%%%%%%%%%%%%%%%%%%%%%%%%%%%%%
\section{Results and data analysis}\label{SectionResults}
%%%%%%%%%%%%%%%%%%%%%%%%%%%%%%%%%%%%%%%%%%%%%%%%%%%%%%%%%%%%%%%%%%%%%%%%

\subsection{Autocorrelation and error analysis}

For all individual runs we determined the integrated autocorrelation for
the observable $E$,
\begin{equation}
\tau_{int,E} = \frac{1}{2} + \sum_{n>0}
{\langle E_0 E_n\rangle - \langle E\rangle^2
\over \langle E^2\rangle - \langle E\rangle^2 }  .
\end{equation}
(Here the index indicates the n-th configuration measured in the
Markov process.)  The inverse value provided us with a weight factor
of the corresponding data sample in the multihistogram analysis.

% Insert figure 1 here
\begin{figure}[htp]
\begin{center}
\epsfig{file=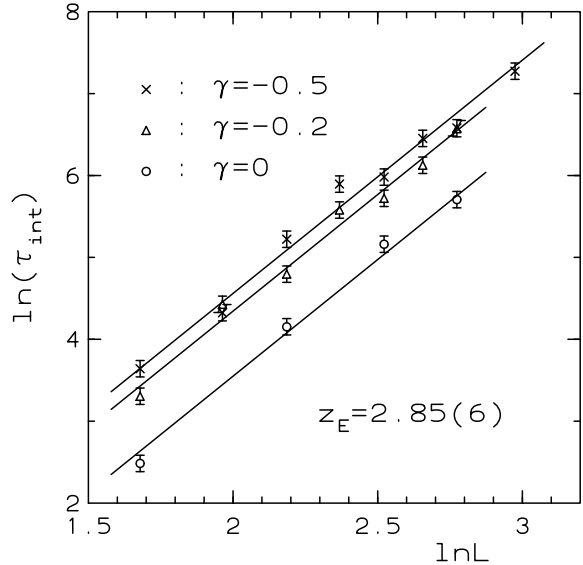,width=8cm}
\end{center}
\caption{\label{tintfitfig}
The maximum values of the integrated autocorrelation length for the
energy observable together with the fit to an exponential dependence
on the lattice size. }
\end{figure}

For the maximum values of the autocorrelation lengths we produced two
sets of values. One was the maximal observed values of
$\tau_{int,E}$ for all samples for
the given lattice size. The other resulted from a fit to the values of
$\tau_{int}(\beta)$ to a peak shaped curve which has its peak position
where the specific-heat (see the discussion below of the analysis
for the cumulants) assumes its maximum. In the subsequent analysis we
discuss only the results due to the first set. The second set led to
similar results.

The maximum values of $\tau_{int,E}$ (cf. Table
\ref{table2}) increase from values of $\simeq 12$ for
$\gamma=0, L=4$ up to $\simeq 1600$ for $\gamma=-0.5, L=12$. This
demonstrates the necessity to work with large samples of several
$10^5$ configurations for each value of $\beta$, at least for the
large lattices.

At second order PTs the maximum
values of the integrated autocorrelation time provides an estimate for
the corresponding dynamical critical exponent $z_E$ through
\begin{equation}\label{tintscaling}
\tau_{int,E}\simeq \min (L,\xi)^{z_E}
\end{equation}
(where $\xi$ denotes the correlation length). At $1^{\rm st}$  order
transitions
one expects that the autocorrelation length grows exponentially
$\simeq \exp{(cL^{D-1})}$.
In Fig.\ref{tintfitfig} a log-log plot shows that the
size dependence is indeed compatible with (\ref{tintscaling}).  We
assume, that the peak values correspond to the point where
\begin{equation}
\tau_{int,E}(\beta_{peak})= c(\gamma) L^{z_E} .
\end{equation}
A simultaneous fit to all three data sets (for the three values of
$\gamma$) gives for the dynamical critical exponent $z_E = 2.85(6)$.
The coefficients grow from $c(0)=0.12$ up to $c(-0.5)=0.32$.  The
results for $\gamma=0$ were obtained with an additional overrelaxation
step in the Metropolis updating.  Although the absolute value of the
autocorrelation lengths decreased by a factor of about 2, the dynamical
critical exponent appears not to be affected.

That value is substantially larger than the value $z=2$ expected for
the random walk dynamics of local algorithms and demonstrated for
Gaussian models. This behavior is indicative of a more complex dynamics
than it is usually anticipated for systems with local excitations. The
non-locality of the monopole loops may be responsible for the observed
effect.  On the other hand, we may not yet be asymptotic and the
determination of reliable values for this exponent is notoriously
difficult.

As a consistency check we also determined autocorrelation times from a
fit to an exponential decay and from blocking analysis. The resulting
values were typically proportional to those discussed above, although
less reliable, i.e. with large statistical fluctuations. The
exponential autocorrelation time and its dynamical critical exponent
are upper bounds to the integrated autocorrelation time (cf.
\cite{So91}).

The statistical errors for all our raw data (i.e. positions and values
of cumulant extrema and positions of the Fisher zeros) were determined
with the jackknife method. From the original set of values $E$ for each
configuration we chose 10 different subsets by omitting 10\% of the
numbers, providing 10 histograms.  The Ferrenberg-Swendsen analysis
then was repeated for all these subhistograms and parameters for the
cumulants (peak positions, values, Fisher zeros) were determined.  The
distribution of these numbers defined the errors according to the
jackknife procedure. The central values were taken from the analysis of
the complete data.  The fits were performed using these central values
and errors.

The simulations on the CRAY-YMP have been performed employing a
vectorized version of the shift-register random number generator, which
in its actual implementation uses XOR operations in between the $i$ and
$i+103$ element to generate the $i+205$ element of the sequence.  For
the programs on the workstations we used a corrected version of
RCARRY \cite{Ja90} based on the ``subtract-and-borrow''
version of a lagged Fibonacci algorithm.

\subsection{Results: data and fits}

We analyzed the final numbers for the pseudocritical points (the
extrema positions of the cumulants and the real part of the position of
the closest Fisher zero), the extrema values of the cumulants and the
imaginary part of the Fisher zero. The fits were performed both, for
all lattices sizes and for a subset of lattices with $N\geq6$, in order
to estimate to which amount we see asymptotic behavior.  

\subsubsection{Histograms}

% Insert figure 2 here
\begin{figure}[htp]
\begin{center}
\epsfig{file=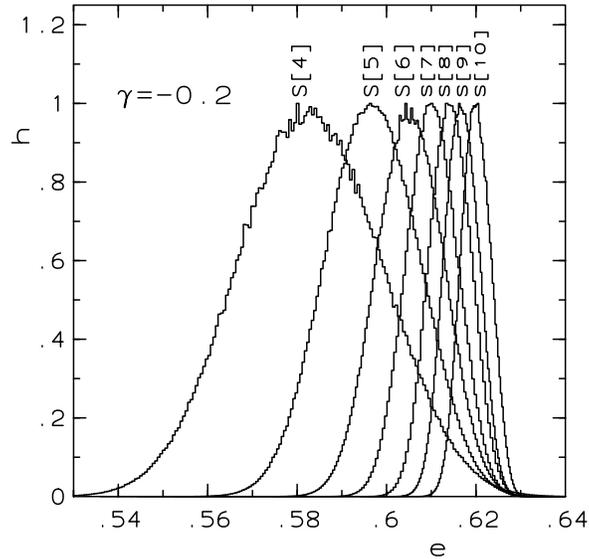,width=8cm}
\end{center}
\caption{\label{FigMFSHistall2} Pseudo-histograms as function of $e$ at
$\gamma=-0.2$ and for lattices size $N=$4, 5, 6, 7, 8, 9, 10 (from left
to right) at the respective peak positions of the specific-heat (in
Table \ref{table3} ) and normalized to unit maximum value.}
\end{figure}

{}From the combination of histograms determined for different values of
$\beta$ according to the Ferrenberg-Swendsen technique we obtain
the distribution densities $\rho(E;\gamma)$ in (\ref{PartFcn}).
A necessary condition for the effectiveness of the approach is sufficient
overlap between the individual histograms. From the densities
one may construct pseudo-histograms (or reweighted histograms)
at arbitrary values of $\beta$ (which should be in the domain covered
by the individual histograms),
\begin{equation}
h(E;\gamma) = \rho(E;\gamma) \exp{(-\beta E)} .
\end{equation}
These interpolate the individual histograms but they also bring 
together and represent all histogram data.

% Insert figure 3 here
\begin{figure}[htp]
\begin{center}
\epsfig{file=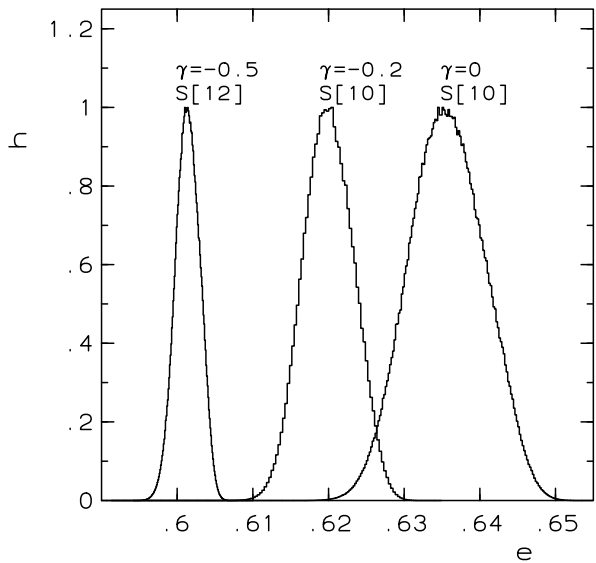,width=8cm}
\end{center}
\caption{\label{FigMFSHist025} Pseudo-histograms vs $e$ for lattices
(from left to right) ${S\,}$[12] ($\gamma=-0.5$), ${S\,}$[10]
($\gamma=-0.2$) and ${S\,}$[10] ($\gamma=0$) at their respective peak
positions of the specific-heat and normalized to unit maximum.}
\end{figure}

In Fig. \ref{FigMFSHistall2} we plot the pseudo-histograms for
$\gamma=-0.2$ for all lattice sizes studied and determined at the peak
positions of the specific-heat. They are normalized to unity at their
respective maxima. No double peak structure is observable.  In Fig.
\ref{FigMFSHist025} the pseudo-histograms for the largest lattice sizes
are plotted for all three values of $\gamma$. (An individual
histogram at $\gamma = 0$ is shown in \cite{JeLa96a}.) Again, there is
no indication of a discontinuity signal.  Such an observation was made
already in the study of the SH lattice (at $\gamma=0$)
in \cite{LaNe}.

Also the individual histograms show no two-peak structure.  Actually,
if the statistics is small, spurious signals may appear, but with
increasing statistics they always vanished.  (They also were not at
consistent positions.) Finally we mention, that there are studies at
established but weak first order transitions (the 2D 5-state Potts
model) on similar spherelike lattices, where a two-state signal has
been observed clearly \cite{HoJa96}. Therefore we find no indication,
that the particular kind of lattice studied here has a tendency to
smear out weak two-state signals.

A two-peak distribution is an indicator of a possible $1^{\rm st}$
order transition. However, in order to establish this order one should
also find further signals for coexistence of phases like FSS consistent
with $\nu=1/D=1/4$, correct scaling of the minimum between the peaks
(suppressed due to the 3D interface) in the distribution and tunneling
probability $\propto \exp{(-2 \sigma L^3)}$. Up to now no consistent
observations of that kind have been made in the U(1) theory for the
toroidal lattices, where one finds two-state signals.

Within the scope of lattice sizes studied here we are therefore led to
assume a $2^{\rm nd}$  order behavior for $\gamma\leq 0$. The
subsequent scaling analysis leads to results fully supporting this
assumption.

In Tables \ref{table3} and \ref{table4} we summarize our
results for the extrema values and positions of the cumulants and of
the positions of the closest Fisher zeros. The analysis of these data
is discussed in the subsequent sections.

%Insert table 3 here
\begin{table}[htp]
\caption{Extrema positions of the cumulants and the
real part of the positions of the closest Fisher zeros.
\label{table3}}
\begin{center}
\begin{tabular}{rrllll}
\hline
$\gamma$&$N$ &$\beta(c_V)$&     $\beta(V_{CLB})$&       $\beta(U_4)$&
        $\mbox{Re}(z_0)$\\
\hline
\hline
0
&4      &1.0027(3)      &0.9990(4)      &1.0051(3)      &1.0047(4)\\
&6      &1.0151(2)      &1.0148(1)      &1.0156(2)      &1.0156(2)\\
&8      &1.0179(1)      &1.0179(1)      &1.0182(1)      &1.0182(1)\\
&10     &1.0183(1)      &1.0183(1)      &1.0185(1)      &1.0185(1)\\
\hline
-0.2
&4      &1.1473(6)      &1.1422(5)      &1.1512(11)     &1.1514(12)\\
&5      &1.1588(4)      &1.1574(5)      &1.1607(5)      &1.1608(5)\\
&6      &1.1640(3)      &1.1634(3)      &1.1652(7)      &1.1650(12)\\
&7      &1.1664(4)      &1.1662(4)      &1.1675(4)      &1.1677(2)\\
&8      &1.1681(1)      &1.1680(1)      &1.1685(3)      &1.1684(3)\\
&9      &1.1688(1)      &1.1687(1)      &1.1690(1)      &1.1690(1)\\
&10     &1.1695(1)      &1.1695(1)      &1.1698(2)      &1.1697(2)\\
\hline
-0.5
&4      &1.4067(7)      &1.3987(10)     &1.4126(15)     &1.4070(42)\\
&5      &1.4202(7)      &1.4177(8)      &1.4239(13)     &1.4246(17)\\
&6      &1.4270(6)      &1.4262(6)      &1.4291(5)      &1.4292(7)\\
&7      &1.4307(4)      &1.4304(4)      &1.4320(6)      &1.4318(5)\\
&8      &1.4325(2)      &1.4324(2)      &1.4328(3)      &1.4328(3)\\
&9      &1.4340(7)      &1.4339(8)      &1.4354(22)     &1.4353(4)\\
&10     &1.4346(2)      &1.4345(2)      &1.4349(2)      &1.4349(2)\\
&12     &1.4359(6)      &1.4359(6)      &1.4366(1)      &1.4365(1)\\
\hline
\end{tabular}
\end{center}
\end{table}

%Insert table 4 here
\begin{table}[htp]
\caption{Extrema values of the cumulants and the
imaginary part of the positions of the closest Fisher
zeros.
\label{table4}}
\begin{center}
\begin{tabular}{lrllll}
\hline
$\gamma$&$N$&   $c_V$&  $V_{CLB}$&      $U_4$&  $\mbox{Im}(z_0)$\\
\hline
\hline
0
&4      &1.85(1)        &-0.142(1)E-02  &2.77(1)        &0.0300(5)\\
&6      &3.93(5)        &-0.362(5)E-03  &2.67(2)        &0.0066(2)\\
&8      &6.75(14)       &-0.157(3)E-03  &2.61(3)        &0.0024(1)\\
&10     &9.47(23)       &-0.792(20)E-04 &2.65(2)        &0.0013(1)\\
\hline
-0.2
&4      &1.22(1)        &-0.982(6)E-03  &2.83(1)        &0.0388(10)\\
&5      &1.63(1)        &-0.395(3)E-03  &2.81(1)        &0.0185(4)\\
&6      &2.07(3)        &-0.201(3)E-03  &2.82(2)        &0.0112(14)\\
&7      &2.54(5)        &-0.117(2)E-03  &2.80(4)        &0.0063(3)\\
&8      &3.08(4)        &-0.755(9)E-04  &2.80(2)        &0.0043(2)\\
&9      &3.54(9)        &-0.503(12)E-04 &2.75(5)        &0.0029(2)\\
&10     &4.22(13)       &-0.371(11)E-04 &2.71(8)        &0.0020(2)\\
\hline
-0.5
&4      &0.76(1)        &-0.647(3)E-03  &2.89(1)        &0.0578(32)\\
&5      &0.95(1)        &-0.246(3)E-03  &2.88(1)        &0.0271(11)\\
&6      &1.16(1)        &-0.121(1)E-03  &2.85(2)        &0.0150(11)\\
&7      &1.36(1)        &-0.673(8)E-04  &2.83(1)        &0.0087(2)\\
&8      &1.62(6)        &-0.427(16)E-04 &2.72(6)        &0.0053(4)\\
&9      &1.69(5)        &-0.258(8)E-04  &2.87(6)        &0.0046(5)\\
&10     &1.98(8)        &-0.188(8)E-04  &2.78(5)        &0.0030(2)\\
&12     &2.26(12)       &-0.946(49)E-05 &2.78(12)       &0.0019(2)\\
\hline
\end{tabular}
\end{center}
\end{table}

\subsubsection{Fisher zeros\label{ResFisherZero}}

The results for the imaginary part of the position $z_0$ of the Fisher
zero closest to the $\beta$ axis are given in Table
\ref{table5}. Although we tried fits including further background
contributions it turned out that the form (\ref{FSSIm}) is sufficient.

% Insert figure 4 here
\begin{figure}[htp]
\begin{center}
\epsfig{file=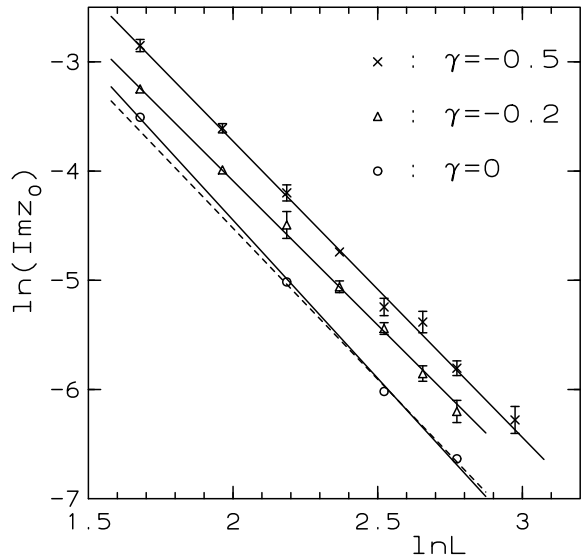,width=8cm}
\end{center}
\caption{\label{FigImz0Fit} Log-log Plot for $\mbox{Im}\,z_0$ vs $L$
with individual fits for each $\gamma$.  At $\gamma = 0$ the full line
denotes our fit to all $N$, the dotted line a fit to data for $N \ge
6$.  For $\gamma < 0$ no visible difference would be seen.}
\end{figure}

%Insert table 5 here
\begin{table}[htp]
\caption{Results for $\nu$ from individual fits to
\protect{$\mbox{Im}\,z_0$ }
according to (\protect{\ref{FSSIm}}).\label{table5}}
\begin{center}
\begin{tabular}{rrr}
\hline
$\gamma$ & $\nu$ & $\chi^2/{d.f.}$\\
\hline
\hline
0       & 0.345(3)      & 4.7\\
-0.2    & 0.378(7)      & 0.3\\
-0.5    & 0.368(8)      & 0.8\\
\hline
\end{tabular}
\end{center}
\end{table}

In Fig.~\ref{FigImz0Fit} we show individual fits for each $\gamma$. For
$\gamma=0$ we distinguish two lines: the fit to all $N$ and one to data
for $N \ge 6$.  In particular for $\gamma=0$ the $N=4$ data seems to be
outside the overall behavior, indicating that at this lattice size the
asymptotic behavior is not yet seen.  According to our interpretation,
we expect the value $\gamma=0$ to be closest to a tricritical point,
which may explain the larger deviations as compared to the other values
of $\gamma$.

A joint fit to the data for all three $\gamma$-values with universal
$\nu$ but individual proportionality factors gives $\nu=$ 0.354(3) at a
$\chi^2/{d.f.}$ value of $2.7$; including only data with $N\geq 5$ we
obtain $\nu=$ 0.368(5) ($\chi^2/{d.f.} =$ 0.98). Finally if we restrict the
fit to the data with $N\geq 6$ we find $\nu=$ 0.365(8) at
a ($\chi^2/d.f=$1.05).  This last fit we consider to be the most
reliable determination of $\nu$ (a corresponding plot may be found in
\cite{JeLa96a}).

\begin{figure}[htp]
\begin{center}
\epsfig{file=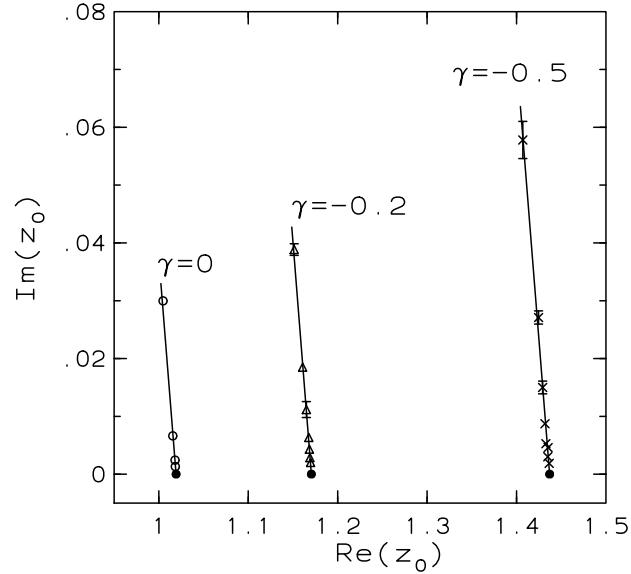,width=8cm}
\end{center}
\caption{Positions of the closest Fisher zeros for all lattice sizes and
all $\gamma$.\label{FigZeroPos} }
\end{figure}

It is interesting to compare the absolute positions of the zeros for
different values of $\gamma$ in Fig. \ref{FigZeroPos}. We find that the
zeros are generally closer to the real axis for $\gamma$ closer to 0.
This indicates, that asymptotic scaling sets in somewhat later (at
larger lattices) at more negative values of $\gamma$. This correlates
with  the peak values of the specific-heat, as will be discussed below
in the discussion of the cumulants.

The results for the real parts of the Fisher zero positions will be
discussed together with the pseudocritical values.

% Insert figure 5 here

\subsubsection{Cumulant values}

\begin{figure}[htp]
\begin{center}
\epsfig{file=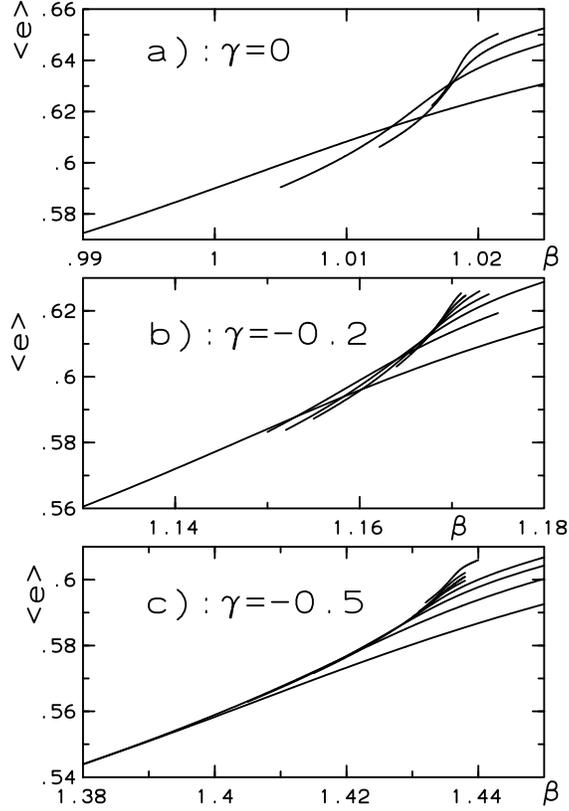,width=10cm}
\end{center}
\caption{Values of $\langle e \rangle$ vs $\beta$ around the
pseudocritical points for the three values of $\gamma$ and
for all lattice sizes:
(a) $\gamma=0$, $N=4,6,8,10$;
(b) $\gamma=-0.2$, $N=4,5,6,7,8,9,10$;
(c) $\gamma=-0.5$, $N=4,5,6,7,8,9,10,12$.
\label{FigEnergy}}
\end{figure}

\begin{figure}[htp]
\begin{center}
\epsfig{file=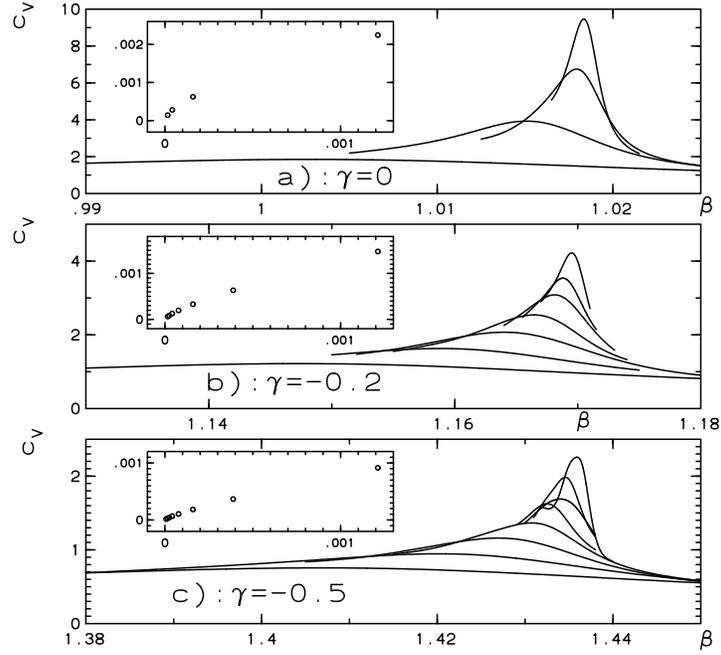,width=8cm}
\end{center}
\caption{Values of $c_V$ vs $\beta$ around the pseudocritical points for
the considered values of $\gamma$ and for all lattice sizes:
(a) $\gamma=0$, $N=4,6,8,10$;
(b) $\gamma=-0.2$, $N=4,5,6,7,8,9,10$;
(c) $\gamma=-0.5$, $N=4,5,6,7,8,9,10,12$.
The inserts exhibit the peak values $c_V/V$ vs $1/V$ demonstrating
their approach towards 0 for $V\to\infty$.
\label{FigCV}}
\end{figure}

\begin{figure}[htp]
\begin{center}
\epsfig{file=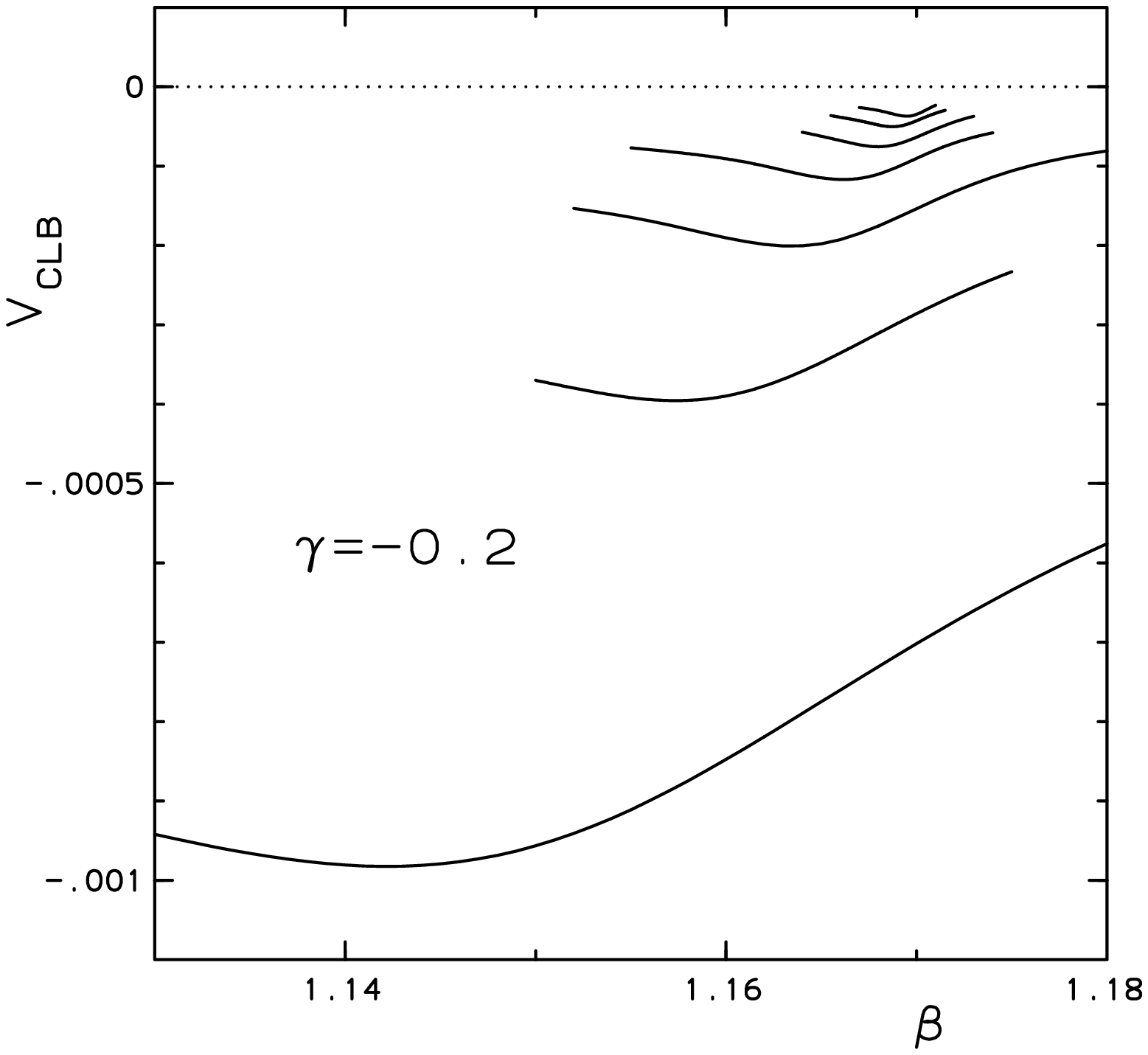,width=8cm}
\end{center}
\caption{$V_{CLB}$ vs $\beta$ for all lattices sizes
studied at $\gamma=-0.2$. For $1^{\rm st}$  order PTs the values
at the minima should asymptotically approach a nonzero
constant. \label{FigVCLB}}
\end{figure}

Figures~\ref{FigEnergy} and \ref{FigCV} exhibit $\langle e \rangle$ and
$c_V$ in the pseudocritical range, and Fig. \ref{FigVCLB} gives an
example for the behavior of $V_{CLB}$. The inserts in Fig. \ref{FigCV}
demonstrate, that the peak values of the specific-heat grow slower than
the volume and that $c_V / V$ approaches zero in the thermodynamic
limit, indicating a $2^{\rm nd}$ order PT.

% Insert figure 6 here

% Insert figure 7 here

% Insert figure 8 here

Our ansatz 
\begin{equation} \label{FitFunctionCV}
c_{V,max}(L)=  a(\gamma)+ b(\gamma) L^{2/\nu(\gamma) - 4}
\end{equation}
for the scaling behavior (\ref{FSSrelCV}) (with Josephson's law
relating $\alpha$ with $\nu$) allows for a background constant.  We
performed various fits restricting the coefficients in different ways.
It turned out, that one should not omit the background constant
$a(\gamma)$. If one does, then the fits become size dependent and have
worse $\chi^2$ if all lattice sizes are included and better if one
omits the small lattices.  We therefore allow for such a background
parameter and include all lattices sizes in the fits.

If we leave $a$, $b$, and $\nu$ $\gamma-$dependent we get consistent
results with $\nu$ varying between 0.361 and 0.404 (cf. Table
\ref{table6}).  If we enforce a $\gamma$-independent value of
$\nu$ we find $\nu=0.378(4)$ and the reasonable $\chi^2=2.2$ but
different values for the background parameters.

%Insert table 6 here
\begin{table}
\caption{Results of the fit to $c_V$ according to (\ref{FitFunctionCV}).
\label{table6}}
\begin{center}
\begin{tabular}{rllll}
\hline
$\gamma$        &$\nu$  &$a(\gamma)$    &$b(\gamma)$
        &$\chi^2/{d.f.}$\\
\hline
\hline
0       &0.361(6)       &0.07(18)       &0.136(34)      &2.3\\
-0.2    &0.374(6)       &0.35(9)        &0.090(20)      &0.3\\
-0.5    &0.404(9)       &0.10(10)       &0.132(44)      &1.0\\
\hline
\end{tabular}
\end{center}
\end{table}

% Insert figure 9 here
\begin{figure}[htp]
\begin{center}
\epsfig{file=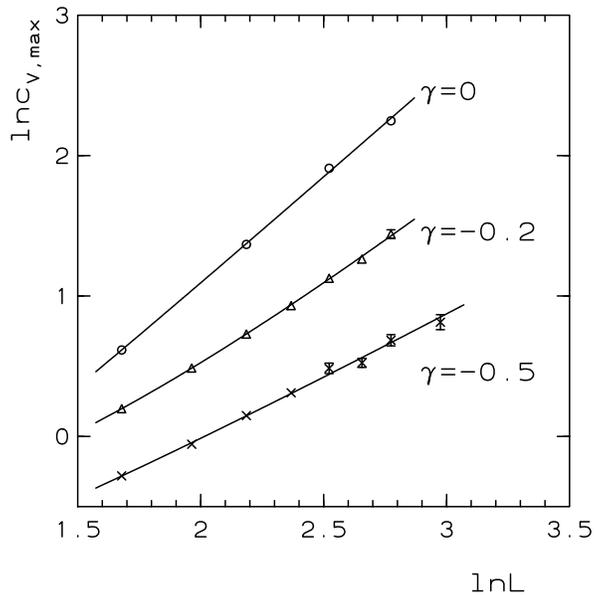,width=8cm}
\end{center}
\caption{This is a log-log plot of specific-heat $c_V$ vs $L$ 
together with the fit results to form
(\ref{FitFunctionCV}) and parameter values from
Table \ref{table6}. \label{FigFitCVValues} }
\end{figure}

Fig.~\ref{FigFitCVValues} is a log-log plot for $c_V$ and the results
of the fits (Table \ref{table6}).  The increase of the value
for $\nu$ with decreasing $\gamma$ indicates that the behavior of $c_V$
is not yet asymptotic. We observed already in the discussion of the
Fisher zero that scaling appears to be retarded towards more negative
values of $\gamma$.  Below (in \ref{SectScalCons}) we try to correct
for this fact by introducing a phenomenological scaling variable.
Indeed we find a consistent scaling behavior of the specific-heat
maximum corresponding to a value of $\nu$ as determined in sect.
\ref{ResFisherZero}.

% Insert figure 10 here
\begin{figure}[htp]
\begin{center}
\epsfig{file=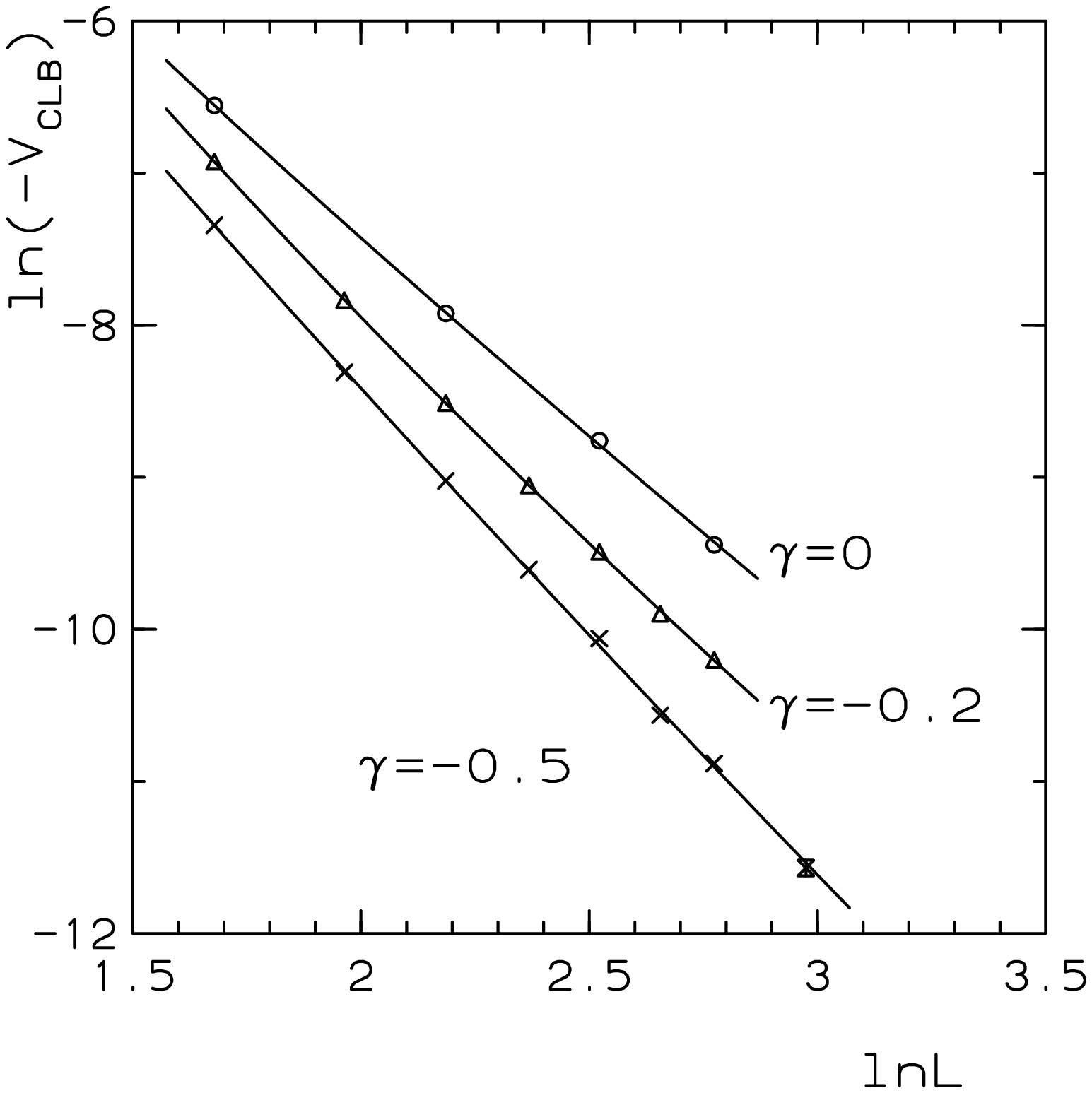,width=8cm}
\end{center}
\caption{A plot of $\ln{(- V_{CLB})}$ vs $\ln L$ for all three
value of $\gamma$, comparing with the fit according
to (\ref{FitFunctionVCLB}).\label{FigFitVCLB} }
\end{figure}

For the CLB-cumulant we found that a fit to the FSS behavior in the form
\begin{equation} \label{FitFunctionVCLB}
V_{ CLB,min }(L)=
\left( a(\gamma) +b(\gamma) L^{2/\nu(\gamma) -4} \right) L^{-4}
\end{equation}
-- in the spirit of the correction term in the specific-heat
(\ref{FitFunctionCV}) -- appears to be suitable.  Fig. \ref{FigFitVCLB}
and Table \ref{table7} show our result. The consistency with the
results for $c_V$ is remarkable.

The values of $V_{CLB}$ clearly tend to vanish in the thermodynamic
limit as expected for $2^{\rm nd}$  order transitions.  It was
unexpected, that the scaling analysis led to sensible results in good
agreement with the results for the specific-heat. (Note that the
CLB-cumulant is a $4^{\rm th}$ order moment and therefore in principle
much more error-prune. For the  -- in comparison to spin model
simulations -- low statistics one cannot put too much confidence in
this quantity.)

%Insert table 7 here
\begin{table}[htp]
\caption{Results for the fit of $V_{ CLB,min }$
according (\ref{FitFunctionVCLB}).
\label{table7} }
\begin{center}
\begin{tabular}{rllll}
\hline
$\gamma$        &$\nu$  &$a(\gamma)$    &$b(\gamma)$
        &$\chi^2/{d.f.}$\\
\hline
\hline
0       &0.361(6)       &-0.23(1)       &-0.071(19)     &2.3\\
-0.2    &0.365(6)       &-0.41(5)       &-0.034(9)      &0.3\\
-0.5    &0.396(9)       &-0.22(5)       &-0.054(19)     &1.1\\
\hline
\end{tabular}
\end{center}
\end{table}

If we omit lattice sizes $N< 6$  the $\chi^2$ improves, but the values
of $\nu$ do not change much. Like for the specific-heat we notice an
increase of $\nu$ fit-values with more negative $\gamma$ which we
interpret as due to the retarded FSS.

The data for $U_4$ show too little size dependence (or have too large
errors) to produce a trustworthy fit to the expected leading scaling
behavior (\ref{FSSrelU4}),
\begin{equation}
U_{4,min}(L)  = a(\gamma) + b(\gamma) L^{4-2/\nu}.
\end{equation}
A joint fit to all data ($\chi^2/{d.f.}$=1.1) gives $\nu=0.35(3)$ and
values of $a= 2.60(3), 2.79(2), 2.81(2)$ which are, however, clearly
different from the value 1 expected at a $1^{\rm st}$  order PT.

\subsubsection{Pseudocritical values}

Let us denote our four definitions for pseudocritical values by
$\beta_c^{(i)}(L)$ (where $i=1\ldots 4$ stands for the peak
positions of $c_V$, $V_{CLB}$ and $U_4$ and $\mbox{Re}\,z_0$,
respectively).
In the fits we allow for the form
\begin{equation} \label{Fitbetapseudo}
\beta_c^{(i)}(L) = \beta_c + a_i L^{-\lambda}.
\end{equation}
For each $\gamma$ we fit simultaneously to all types $i$ for a unique
$\beta_c$ and $\lambda$ but individual values $a_i$.  We find that
allowing for another term $O(L^{-2})$ --- as it is motivated from
the possible contribution of the curvature or lattice inhomogeneities
and as it seemed to be necessary for the
analysis of the SH results in \cite{LaNe} --- does not improve the
$\chi^2$ significantly.

%Insert table 8 here
\begin{table}[htp]
\caption{Results of the fits to the finite-size dependence
of the four definitions of pseudocritical points according to
(\protect{\ref{Fitbetapseudo}}).\label{table8}}
\begin{center}
\begin{tabular}{rlll}
\hline
$\gamma$        &$\beta_c$      &$1/\lambda$    &$\chi^2/{d.f.}$\\
\hline
\hline
0.      &1.0190(1)      &0.321(7)       &5.8\\
-0.2    &1.1709(2)      &0.386(10)      &1.6\\
-0.5    &1.4381(1)      &0.472(12)      &2.5\\
\hline
\end{tabular}
\end{center}
\end{table}

% Insert figure 11 here
\begin{figure}[htp]
\begin{center}
\epsfig{file=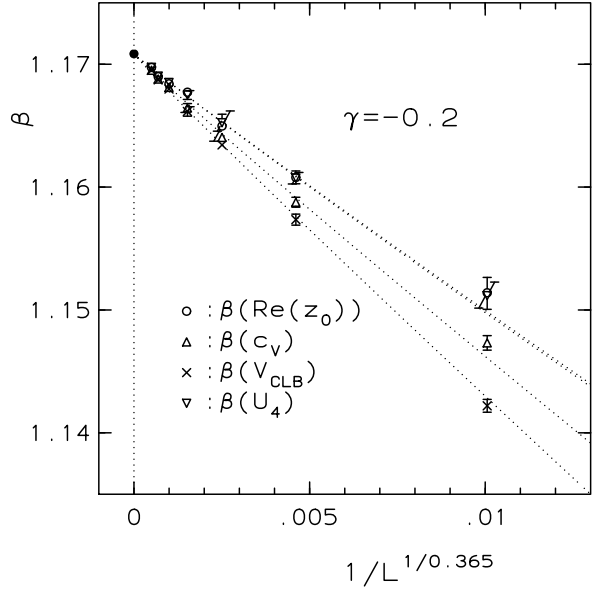,width=8cm}
\end{center}
\caption{
Fits to the data (for $\gamma=-0.2$) for the pseudocritical points
(according to the four different definitions described). We use the
abscissa variable $L^{-1/\nu}$ (for the preferred value $\nu=0.365$)
in order to emphasize the nontrivial dependence.
\label{FigFitbetaps}}
\end{figure}

Table \ref{table8} gives the fit values for the pseudocritical points
and $1/\lambda$. It is not generally true, that $\lambda=1/\nu$ (cf.
the discussion in \cite{Ba83}) and indeed a recent study indicated a
different value for the 2D Ising model on spherelike lattices
\cite{HoLa96}.  Accepting this {\em caveat} we still find numbers of
similar size.  If we allow for a correction term $O(L^{-2})$ due to the
background curvature of our lattices and fix $\lambda=1/0.37$ (i.e. at
a value $1/\nu$ suggested from the other data) the fit is of comparable
quality with compatible values for $\beta_c(\gamma)$ and the fit curves
in the plots are indistinguishable by eye.

Altogether the errors on the pseudocritical points are larger but the
fits are not very satisfying (cf. Fig.  \ref{FigFitbetaps} as an
example; the data and fits for the other $\gamma$-values look
similar).  The value of $\lambda$ is not stringently determined by the
data (or the theory).

\subsubsection{Scaling consistency}\label{SectScalCons}

At finite lattices there are always corrections to FSS, depending on
size, geometry, topology and of course details of the action and the
observables. Since $\mbox{Im}\,z_0$ gives the cleanest FSS signal, we
use it as a phenomenological scaling variable
\begin{equation} \label{DefineSVx}
x \equiv \mbox{Im}\,z_0
\end{equation}
and study in this section the other observables as functions of
$x$.  (Notice that $x$ is defined from the data!) That provides us on
one hand with a consistency check for our results. On the other hand
this assumption allows us to bring together and combine results from
different values of $\gamma$.
% Insert figure 12 here
\begin{figure}[htp]
\begin{center}
\epsfig{file=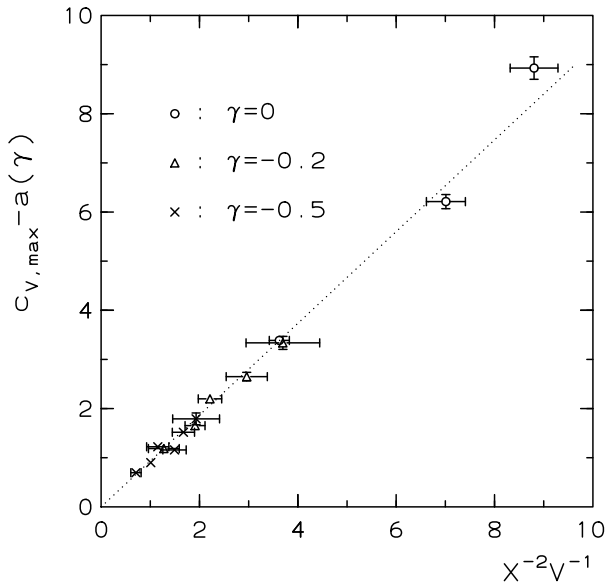,width=8cm}
\end{center}
\caption{A plot of the peak values of the specific-heat (all $\gamma$, all
lattices sizes) vs $1 /(x^2 L^4)$ ($x$ is the phenomenological
scaling variable defined in (\ref{DefineSVx}); the corresponding error bars
are shown as well.
The line represents a linear fit. \label{FigSpecHeatvsx} }
\end{figure}

Assuming the FSS relation $x\simeq L^{-1/\nu}$ (i.e. for
$\mbox{Im}\,z_0$) we expect e.g. for the specific-heat the behavior
(cf. (\ref{FSSrelCV}))
\begin{equation}
c_{V,max}(L)    \simeq a + L^{2/\nu-4} \simeq a + b x^{-2} L^{-4} ,
\end{equation}
where we -- as discussed -- allow for the additive constant to
represent the unknown background. Fig.~\ref{FigSpecHeatvsx} shows the
peak values of the specific-heat for the data vs $x^{-2} L^{-4}$
together with the result of a linear fit to the data for $N\geq 5$.
For this fit we had to assume different values of the background
constant $a=0.22, 0.55, 0.40$ for different $\gamma=0, -0.2, -0.5$.
The plot demonstrates the consistency of the FSS of the specific-heat
values with that of $\mbox{Im}\,z_0$.  This is not surprising, since
these quantities have a close relationship as discussed in sect.
\ref{SecFisherZeros}.

\subsubsection{Summary of the fit results}

The cleanest and most consistent results on FSS come from the imaginary
part of the closest Fisher zero positions and suggest a value
$\nu=0.365(8)$.

The peak values of the various cumulants are generally consistent with
$\nu\simeq 0.35\ldots 0.40$ although they seem to prefer larger values
of $\nu$ towards more negative values of $\gamma$; this may be
explained by later onset of scaling.  Allowing for a smooth background
(in the neighborhood of the peak) like a constant added to the 
specific-heat improves the consistency of the scaling.

Even the pseudocritical positions --- although with less predictive
power due to the uncertainty of the relation of the shift exponent
$\lambda$ to the inverse critical exponent $1/\nu$ --- show scaling
with values of $1/\lambda$ at least roughly of the same magnitude.

In general we find better scaling behavior than for the more ``edgy''
lattices types SH studied in \cite{LaNe}.  Corrections to FSS
due to the curvature (which may be of $O(L^{-2})$) do not seem to be
necessary.

%%%%%%%%%%%%%%%%%%%%%%%%%%%%%%%%%%%%%%%%%%%%%%%%%%%%%%%%%%%%%%%%%%%%%%%%
\section{Conclusion}
%%%%%%%%%%%%%%%%%%%%%%%%%%%%%%%%%%%%%%%%%%%%%%%%%%%%%%%%%%%%%%%%%%%%%%%%

After 17 years \cite{CrJa79b} of development it has now been possible
to obtain a consistent picture of the scaling behavior in the pure
compact U(1) lattice gauge theory at the confinement-Coulomb phase
transition.  Using a radically new kind of finite lattices, modern FSS
methods and substantial computer power we have found a consistent
picture of the scaling behavior of several bulk observables. Within the
limits of numerical evidence, our analysis strongly suggests the
existence of a non-Gaussian fixed point with $\nu$ distinctly different
from 1/2 (Gaussian value) or 1/4 (1$^{\rm st}$ order transition). Its
universality class extends in the $\beta$ -- $\gamma$ plane along the
phase transition line at negative $\gamma$, but includes also the
Wilson action, $\gamma = 0$.

This implies that using RG methods, one can construct a unitary
continuum field theory in 4D which is neither asymptotically free nor
trivial. As we point out in \cite{JeLa96a}, this holds also for several
theories related to the U(1) lattice theory by duality
transformations.  Thus rather than being an exercise ground for lattice
QCD, the pure compact U(1) lattice gauge theory at its phase transition
defines a sort of quantum field theory in 4D which is not used either
in the Standard Model nor in its presently known extensions.

The natural question is whether these novel features of the U(1)
lattice theory might be related to the difficulties encountered in its
numerical investigation. The strongly interacting monopole loops are an
obvious candidate for concern. The previous studies using the surface
of the 5D cubic lattice \cite{LaNe}, as well as our present results,
suggest that the topology of the finite lattice is crucial. For $\gamma
\le 0$ the two-state signal vanishes on lattices with spherelike
topology. However, this fact alone does not yet confirm the
speculations that the winding monopole loops are the culprits. A more
detailed information about the field configurations, and more
experience with various lattices and boundary conditions are required.
It could be that the question is indeed more than technical and its
pursuit might lead to a deeper understanding of the non-Gaussian fixed
point.

\paragraph{Acknowledgments: }
We wish to thank Ch.~Hoelbling and U.-J.~Wiese for discussions. The
computations have been performed in part on the CRAY-YMP of HLRZ
J\"ulich and the Parallel Compute Server of KFU Graz.  The work has
been supported by Deutsches BMBF.

\clearpage
%%%%%%%%%%%%%%%%%%%%%%%%%%%%%%%%%%%%%%%%%%%%%%%%%%%%%%%%%%%%%%%%%%%%%%%%
\section*{Appendix: Lattice details}
%%%%%%%%%%%%%%%%%%%%%%%%%%%%%%%%%%%%%%%%%%%%%%%%%%%%%%%%%%%%%%%%%%%%%%%%

The lattice types SH[N] (the surface of an $N^5$
hypercube) and S[N] have the same link connectivity
structure.  The lattice is build from plaquettes with 4 links each.
Not all sites have 8 links, not all links are bordering 6 plaquettes,
and not all plaquettes are faces of exactly four 3D cubes, as it is the
case for the hypertorus.  All 3D cubes are bordering exactly four 4D
cubes.

Let us denote the total number of sites by $n_s$, and the number of
sites with $i$ links by $n_{s,i}$.  A corresponding notation holds for
the links and plaquettes.  We have
\vspace{12pt}

\begin{tabular}{lll}
sites:  & $n_s$       &$= 10 (N-1)^4 + 20 (N-1)^2 +2$\\
        & $n_{s,5}$   &$=32$\\
        & $n_{s,6}$   &$=80 (N-1)- 80$\\
        & $n_{s,7}$   &$=80(N-1)^2 - 160 (N-1)+80$\\
        & $n_{s,8}$   &$=10 (N-1)^4 - 60 (N-1)^2 $\\
        &               &\phantom{=}$+ 80 (N-1) -30$\\
links:  & $n_l$               &$= 40 (N-1)^4 + 40 (N-1)^2$\\
        & $n_{l,4}$   &$= 80 (N-1)$\\
        & $n_{l,5}$   &$= 160 (N-1)^2 -160 (N-1)$\\
        & $n_{l,6}$   &$= 40 (N-1)^4 -120 (N-1)^2 $\\
        &               &\phantom{=}$+80 (N-1)$\\
plaquettes:& $n_p$             &$= 60 (N-1)^4 + 20 (N-1)^2$\\
        & $n_{p,3}$   &$= 80 (N-1)^2$\\
        & $n_{p,4}$   &$= 60 (N-1)^4 - 60 (N-1)^2$\\
3D cubes:& $n_{3c}$    &$= 40 (N-1)^4$\\
4D cubes:& $n_{4c}$    &$= 10 (N-1)^4$\\
\end{tabular}
\vspace{12pt}

For example, for $N=12$ one has $n_s = 148832 \simeq (19.6^4) \simeq V$.

The number of plaquettes with just  three 3D cubes (as well as the
corresponding numbers for links and sites) is suppressed relative to
the leading terms in $O(1/N^2)$. This is typical for contributions due
to curvature. We may say that the lattice becomes locally flat with
$O(1/N^2)$.

Contrary to the usual hypercubic torus this lattice is not self-dual.
Possible monopole loops live on the dual lattice ${S\!H\,}'$,
which does have a few ($n_{p,3}$) plaquettes of 3 links in addition to
the usual ones.

Euler's relation for spherelike lattices
(of the type discussed, i.e.
without further holes) is
\begin{equation}
n_s-n_l+n_p-n_{3c }+ n_{4c} = 2
\end{equation}
(whereas it is zero for the torus).

The lattice SH --- in analogy to the 2D situation  ---
may be imagined as an ensemble of 10 hypercubic lattices, glued
together at their boundaries. The lattice ${S\,}$ is
constructed by a projection of SH from its center onto
the unit sphere $S^4$.

\end{document}